\documentclass [aps, twocolumn, bibliography,floatfix]{revtex4-2}
\usepackage{graphicx,dcolumn,bm,amssymb,amsmath,hyperref,epsf,xcolor,float}
\usepackage{longtable}

\def\kCuBr{$\kappa$-(BEDT-TTF)$_2$\-Cu[N(CN)$_2$]Br}
\def\kCuCl{$\kappa$-(BEDT-TTF)$_2$\-Cu[N(CN)$_2$]Cl}

\def\cm{cm$^{-1}$}

\begin{document}

\title{Charge-sensitive vibrational modes in BEDT-TTF salts: \\Signatures of charge ordering and site charge}

\author{Savita Priya}
\affiliation{1.~Physikalisches Institut, Universit\"at Stuttgart, Pfaffenwaldring 57, 70569 Stuttgart, Germany}
\author{Martin Dressel}
\affiliation{1.~Physikalisches Institut, Universit\"at Stuttgart, Pfaffenwaldring 57, 70569 Stuttgart, Germany}
\author{Jesse Liebman}
\affiliation{Department of Physics and Astronomy, Johns Hopkins University, Baltimore, MD 21218, U.S.A.}
\author{Natalia Drichko}
\affiliation{Department of Physics and Astronomy, Johns Hopkins University, Baltimore, MD 21218, U.S.A.}

\date{\today}

\begin{abstract}
BEDT-TTF-based organic conductors host a number of ground states, tuned by electron repulsion from Mott and charge ordered insulators to superconductors. Knowing charge distribution on the molecular sites in the insulating state of these materials is a key to understanding the origin of these ground states. We survey and discuss the C=C stretching modes in BEDT-TTF based molecular conductors. These molecular vibrations are extremely crucial in characterization of charge-ordered insulators, and are recently linked to superconductivity in some compounds. Focusing on the known examples of BEDT-TTF$^{+0.5}$ salts, we analyse the reliability of the C=C stretching modes for the determination of charge ordering and absolute site charge. Considering the charge-ordered states, a prominent shift in frequency of 141~\cm{} per elementary charge $e$ for $\nu_{27}(b_{1u})$ and 98~\cm/$e$ for $\nu_2$($a_g$) can be clearly realised, however, the distribution resulting from different compounds span over 20~\cm. For nominal BEDT-TTF$^{+0.5}$ compounds, the distribution of the resonance also extends around 20~\cm, yielding an unexpected large uncertainty of $\Delta\rho~\approx~(~\pm~0.045)e$, which is presumably due to the influence of small differences in the structure. This highlights the limitations of charge-frequency relations to detect small deviations in absolute charge values on molecular lattice sites, and emphasises on the use of the relations to estimate charge-ordering, rather than absolute site charge.
\end{abstract}

\maketitle

\section{Introduction}

Low-dimensional model systems are a cornucopia of quantum phenomena, ever so intriguing for condensed matter scientists and keeping them occupied for decades, experimentally demonstrating intrinsic states like Mott insulator, charge order insulator, unconventional superconductivity, anti-ferromagnetic insulator, quantum spin liquid, quantum dipole liquid, etc., at low temperatures. These states originate from the complex interplay of the foundational interactions in solids like electron-phonon coupling, scattering (electron-electron, electron-phonon, phonon-phonon, Umklapp), disorder, and spin-orbit coupling; along with exotic phenomena including electronic correlations, fluctuations (spin and/or charge), charge transfer, and long-range order magnetic interactions \cite{Morosan12,Dagotto05,Dobrosavljevic12,Sigrist91,Basov11,Powell11,Cross79,Capone10,Kaveh84,Jujo02, Kotliar04} The origin and full understanding of many of these states are still under critical discussion and become extremely complex in the presence of multiple-faceted interplay in strongly correlated classes of heavy fermions, iron-pnictide superconductors, high-\textit{T$_c$} cuprates, and other transition metal compounds.

To that end, one-dimensional (TMTTF or TMTSF based) and subsequently two-dimensional (most commonly BEDT-TTF based) molecular systems offer simplified interactions, due to hybridised $\pi$-orbitals forming the conduction band (and lack \textit{d-}orbitals and \textit{f-}orbitals near the Fermi level) \cite{ToyotaBook,Jerome04,Dressel04,Graja92,Ishiguro90,Kotliar04}; this, however, is achieved at the expense of working with fragile as-grown samples; due to which the methods probing the surface electronic states, including angle-resolved photo\-emission spectroscopy (ARPES) and scanning tunneling microscopy (STM) studies are relatively scarce in the molecular systems, owing to technical challenges \cite{Sing03,Torkzadeh25}. As a result, other spectroscopic methods are employed to explore the electronic, structural, spin, dipole and magnetic states of the spins in the molecular systems.

Optical techniques, namely infrared spectroscopy and Raman scattering spectroscopy, present as a reliable indirect techniques to study the electronic states by analysing not only the electronic continuum related to electronic excitations, but also the molecular vibrations, which registers peculiar electronic response of the charge transfer salts and their frequency is easily accessible in the energy range of infrared and Raman spectroscopy \cite{Dressel04}. In BEDT-TTF based charge transfer salts, the C=C stretching molecular vibrations have been extensively utilised as a way to probe charge distribution on the BEDT-TTF lattice sites in the conducting layer, for charge-ordered insulators and by extension in normal states of some non-stoichiometric systems \cite{Yamamoto05,Ivek11,Hassan18} [BEDT-TTF is abbreviated for bis\-(ethy\-lene\-di\-thio)\-tetrathiafulvalene].  

\begin{figure*}
    \centering
        \includegraphics[width=2\columnwidth]{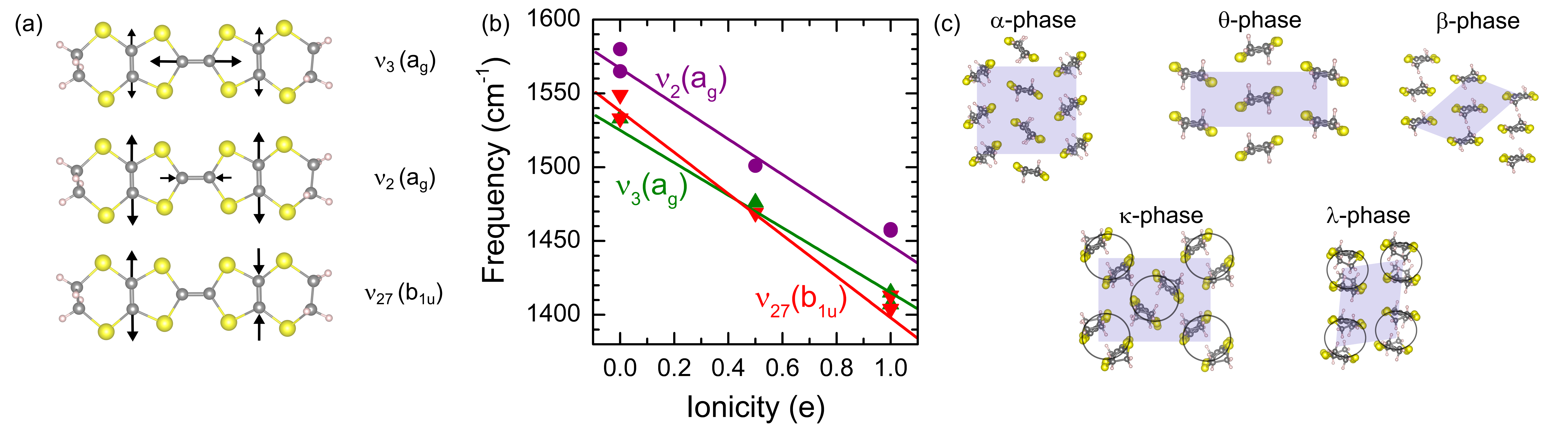}
        \caption{(a) C=C stretching modes for BEDT-TTF molecules: $\nu_2(a_g)$, $\nu_3(a_g)$ and $\nu_{27}(b_{1u})$ (in the notation followed for the to $D_{2h}$ molecular symmetry). (b) Dependence of the C=C stretching modes of the BEDT-TTF molecule on the ionicity. The symbols represent the frequencies calculated by Girlando \cite{Girlando11} while the lines correspond to the regression of the experimental data collected by Yamamoto {\it et al.} \cite{Yamamoto05}. (c) Different stacking patterns found in BEDT-TTF salts: $\alpha$ and $\theta$ have herringbone arrangement, $\beta$ phase has stacked arrangement, $\kappa$ and $\lambda$ phases are formed by dimers, arranged in triangular or stacked patterns, respectively.
        \label{fig:VibrationalModes}
        }
\end{figure*}

Here, we present a review, focusing on the C=C stretching molecular vibrations, namely $\nu_{27}(b_{1u})$, $\nu_2$($a_g$) and $\nu_3$($a_g$) of BEDT-TTF based quasi-two-dimensional charge transfer salts; these notations are presented in the widely accepted $D_{2h}$ symmetry for BEDT-TTF molecules. Out of the aforementioned modes, $\nu_{27}(b_{1u})$ has recently been utilized to drive and identify superconductivity phase, by light-induced or cavity alteration methods, in \kCuBr{} \cite{buzzi2020photomolecular,buzzi2021phase,keren2026cavity}. The C=C stretching modes are crucial as these are the molecular vibrations relatively isolated in energy (observed in mid-infrared region) and not driving any interactions directly (in equilibrium conditions). Instead, these modes act as an important probe for intrinsic electronic response, as the bands near Fermi level, particularly the highest occupied molecular orbital (HOMO) is delocalised over the C=C bonds of the central TTF core \cite{Scriven09a, Scriven09b}. There are several early and recent review articles, focusing on broad and distinct examples of electronic properties of the BEDT-TTF based systems \cite{Dressel04,Seo04,Miyagawa04,Yakushi12,Tomic15,Dressel20,naito2021modern}, which can be referred for a general overview of the class of molecular conductors and the interesting physics accompanying it; but here, we focus solely on the C=C molecular stretching vibrations.

The structure of this review is as follows: Section II gives a brief outline of the C=C vibrations under discussion and the relation between site charge and frequency of the C=C modes, Section III gives examples of charge-ordered states of BEDT-TTF based systems, Section IV is focused on nominal (BEDT-TTF)$^{+0.5}$ compounds in normal state, Section V provides a brief description of $\nu_3$ ($a_g$) mode, and the Section VI presents the summary and conclusions of this review; followed by the tabulated summary of all the vibrational modes (scaled to 300 K) and reported values with appropriate references in Tables II-VII.

\section{Charge transfer and \\C=C stretching vibrations}

Research on molecular conductors accelerated after the discovery of a metallic response in TTF-TCNQ and observation of superconductivity in quasi-one-dimensional TMTSF salts and later in quasi-two-dimensional BEDT-TTF based salts, eventually led to synthesis of several donor-acceptor type systems, forming charge-transfer salts and fullerene, as well as several theoretical studies \cite{naito2021modern}. Some of these works, notably from M. Kozlov, J. Eldridge, A. Girlando, H. H. Wang, K. Yakushi focused on molecular vibrations and its investigation for electron-phonon coupling and electron-molecular vibrational coupling. Through extensive experiments on $^{2}$D or $^{13}$C-isotope containing compounds, as well as high-pressure studies, several key molecular vibrations were identified and accelerated the research of BEDT-TTF based organic charge transfer salts using vibrational studies \cite{Kozlov87,Kozlov89,Maksimuk01,Moldenhauer93b,Kornelsen91,Dressel92,Lin99,Eldridge02}. These are the three C=C stretching vibrations: symmetric stretching of central C=C, $\nu_3(a_g)$ (Raman active), symmetric stretching of C=C bond on the inner rings, $\nu_2(a_g)$ (Raman active), and asymmetric stretching of C=C bond on the inner rings, $\nu_{27}(b_{1u})$ (infrared active), as illustrated in Fig.~\ref{fig:VibrationalModes}(a).

Fig.~\ref{fig:VibrationalModes}(b) summarizes the common understanding: of the three C=C stretching modes, $\nu_{27}(b_{1u})$ infrared active and $\nu_2$($a_g$) Raman active modes are widely utilized for evaluating the charge disproportionation in charge-ordered compounds \cite{Moldenhauer93b,Wojciechowski03,Dressel04,Yamamoto06,Yamamoto08,Olejniczak09,Kaiser10,Yue10,Yamamoto11,Ivek11,Sedlmeier12,Girlando12,Yakushi12,Drichko14,Yakushi15,Tomic15,Beyer16,Ivek17, Hassan18,*Hassan20,Clay19,Mizukoshi20,Kinoshita22,Olejniczak22,Liebman25}, with emphasis on the {\em relative changes} i.e. the slope of the lines in Fig.~\ref{fig:VibrationalModes}(b). The frequencies of these modes follow a linear regression, confirmed experimentally by Yakushi, Yamamoto and collaborators \cite{Yamamoto02,Yamamoto05} and is essentially captured by Girlando's analysis of the vibrational dynamics through first-principle calculations \cite{Girlando11}. Theoretical studies indicate that the $\nu_{3}$($a_g$) mode should behave similar to $\nu_{27}(b_{1u})$, and $\nu_{2}$($a_g$) with change in ionicity of BEDT-TTF molecule, as shown in Fig.~\ref{fig:VibrationalModes}(b); however, Raman scattering results evidence that $\nu_{3}$($a_g$) is not affected by charge, and the 
$\nu_{3}$($b_3g$) is affected by the electron-molecular-vibrational (emv) interaction instead of charge \cite{Moldenhauer93b,Wojciechowski03,Yamamoto06,Yamamoto08,Olejniczak09,Kaiser10,Yue10,Yamamoto11,Yakushi12,Drichko14,Yakushi15,Hassan18,*Hassan20,Clay19,Mizukoshi20,Kinoshita22,Olejniczak22,Liebman25}.

The linear relationship between charge imbalance, $\delta \rho$ and splitting of the molecular vibrational frequency, $\delta\nu$ was first proposed by Mol\-den\-hauer \textit{et al.} \cite{Moldenhauer93b} for \mbox{$\alpha$-(BEDT-TTF)$_2$I$_3$}. H.H. Wang {\it et al}., based on Raman scattering measurements, suggested that the Raman active C=C stretching modes act as the most sensitive probe for the oxidation state of the donor molecules in BEDT-TTF systems \cite{Wang94,Wang96}. The data measured for several salts reveal a spread of several wavenumbers, suggesting simple linear relations for the Raman active modes: $\nu_2(\rho) = 1453~{\rm cm}^{-1} + 86 (1-\rho)~{\rm cm}^{-1}$ and $\nu_3(\rho) = 1420~{\rm cm}^{-1} + 88 (1-\rho)~{\rm cm}^{-1}$. Later, Yakushi's group \cite{Yamamoto05} revised this formula, based on the comparison of a larger number of compounds, particularly focusing on charge-ordered systems and including the $\nu_{27}$ mode, giving the relations: $\nu_2(\rho) = 1447~{\rm cm}^{-1} + 120 (1-\rho)~{\rm cm}^{-1}$ and $\nu_{27}(\rho) = 1398~{\rm cm}^{-1} + 140 (1-\rho)~{\rm cm}^{-1}$. Other fully symmetric $a_g$ modes, such as $\nu_6$, $\nu_9$ and $\nu_{10}$ are less sensitive to the oxidation state of BEDT-TTF \cite{Dressel04, Girlando11}. In the sibling compound BEDO-TTF, very similar relations have been established \cite{Moldenhauer93a,Pokhodnia93,Drozdova00}
\footnote{From Raman scattering experiments on BEDO-TTF molecules \cite{Drozdova00} the relations $\nu_2(\rho) = 1587 + 74.1 (1-\rho)$ and $\nu_3(\rho) = 1416 + 109 (1-\rho)$ were concluded.}.

BEDT-TTF has the tendency to form multiple arrangements in layers on charge transfer salt crystallisation, as illustrated in Fig.~\ref{fig:VibrationalModes}(c), depending on the reaction conditions of electro-crystallization parameters. The different arrangements are denoted by various Greek alphabets: $\alpha$-phase (herringbone arrangement, with two dissimilar stacks), $\theta$-phase (uniform herringbone arrangement), $\beta$-phase (parallel stacks; if stacks are displaced: $\beta''$-phase), $\kappa$-phase (BEDT-TTF dimers are rotated with respect to each other), and $\lambda$-phase (BEDT-TTF dimers are arranged in columnar stacks) \cite{Mori98a, Mori99b, Mori99c}. For $\alpha$-, $\beta$- and $\theta$-phases, the charge described on BEDT-TTF, to balance the monovalent anion gives (BEDT-TTF)$^{+0.5}$, while for dimerized $\kappa$- and $\lambda$-phases, the charge is described as (BEDT-TTF)$_2^{+1}$ [according to charge distribution, this is equivalent to (BEDT-TTF)$^{+0.5}$].

\section{Charge Order}
\label{sec:ChargeOrder}

In a number of BEDT-TTF salts, particularly for the \mbox{$\theta$-phase} and \mbox{$\alpha$-phase} materials, the amount of electrons residing on each molecule is found to be unequal even in normal state \cite{Yakushi01,Dressel04,Yamamoto05,Yamamoto11,Yakushi12,Tomic15,Ivek11}. This charge imbalance can be driven by cryptographically different sites or caused by electronic correlations \cite{Fukuyama00,Seo00,Seo04}. When charge ordered state is established upon cooling, the charge of +1\textit{e} is redistributed on two neighbouring molecules and is accompanied by pronounced changes in various physical properties occur, for instance metal-insulator transitions, dielectric anomalies, second-harmonic generation \cite{Monceau12,Tomic15,Lang25}. Besides x-ray scattering and NMR experiments, vibrational spectroscopy has been established as an extremely sensitive local probe of charge disproportionation.

\begin{figure}[t]
    \centering
        \includegraphics[width=0.81\columnwidth]{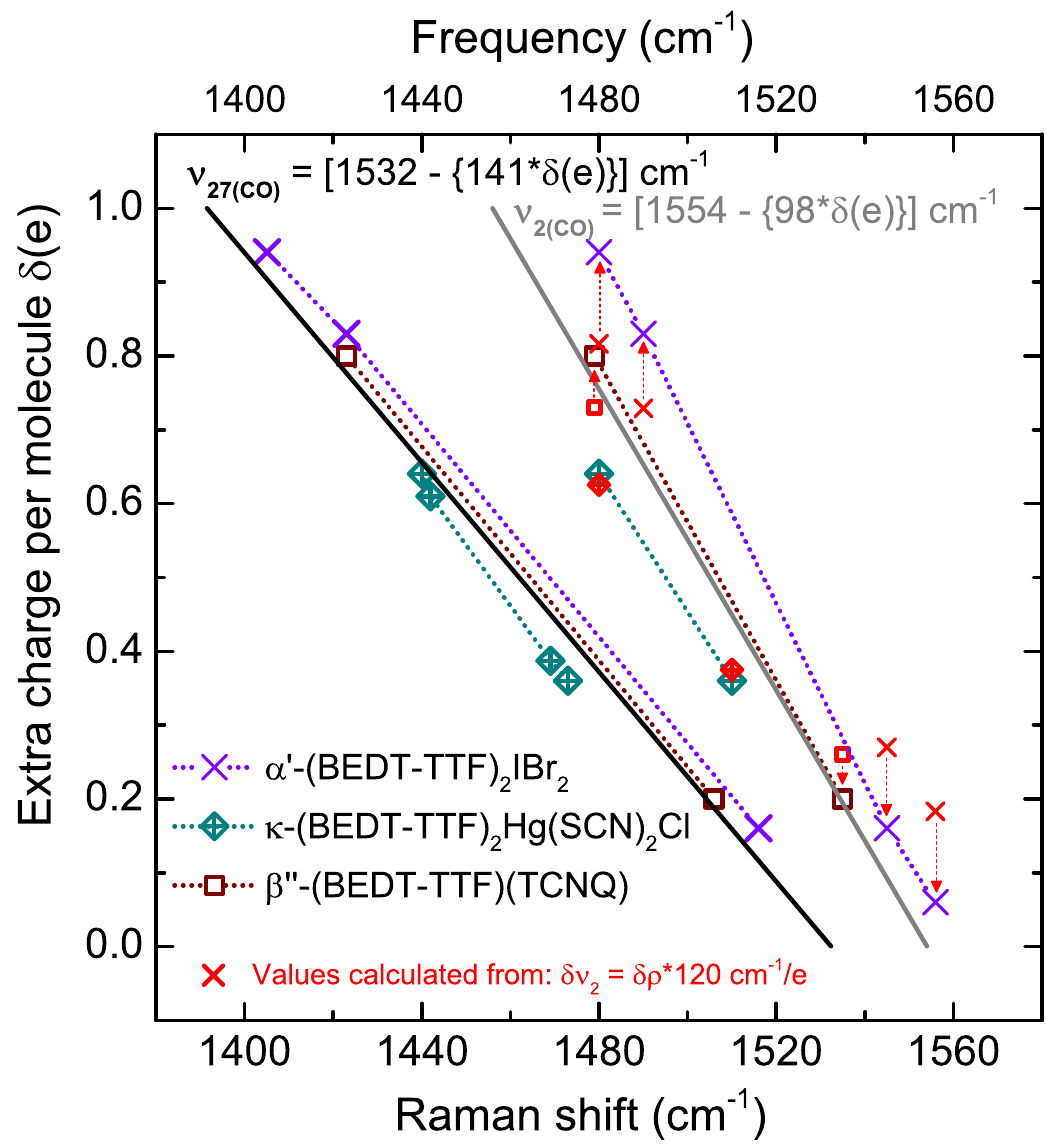}
        \caption{Positions of the $\nu_2(a_g)$ and the $\nu_{27}(b_{1u})$ modes for the BEDT-TTF-based salts for which charge disproportionation was identified by both Raman and infrared spectroscopy (the solid lines indicate linear fits, considering all the examples of charge ordering plotted). The exact compositions and frequencies are given in Table~\ref{tab:splitting}. The red points in the plots are calculated with $\delta\nu_{2}   =  120~\frac{\rm cm^{-1}}{e} \delta\rho$, which are inconsistent with the $ \delta\rho$ values, calculated from $\nu_{27}(b_{1u})$. These values of $\nu_2(a_g)$ are shifted vertically to match $\delta(e)$ values of $\nu_{27}(b_{1u})$.}
    \label{fig:splitting1}
\end{figure}

The relations to find the charge disproportionation by C=C modes are: $\delta\nu_{27}  =  140~\frac{\rm cm^{-1}}{e} \delta\rho$, and $\delta\nu_{2}   =  120~\frac{\rm cm^{-1}}{e} \delta\rho$ \cite{Yakushi12}. Fig.~\ref{fig:splitting1} displays three compounds, where charge-ordered state is observed and the Raman and infrared vibrational modes are reported and are largely symmetric. Here, the relations described in Ref. \cite{Yakushi12}, give different  charge disproportionation values when estimated through $\delta\nu_{27}$ and $\delta\nu_{2}$, highlighting the inconsistency between the relations for the two modes. To make these relations consistent with each other, we shift the charge values of $\nu_2(a_g)$ to match the charge calculated from the frequency values of $\nu_{27}(b_{1u})$ [red points values in Fig. \ref{fig:splitting1} are calculated with: $\delta\nu_{2}   =  120~\frac{\rm cm^{-1}}{e} \delta\rho$, which are shifted vertically to match charge values calculated for $\nu_{27}(b_{1u})$ and have a symmetric splitting]. The new slope for $\nu_2(a_g)$ for charge-ordered state, consistent with $\nu_{27}(b_{1u})$ extends over 98 \cm, when extrapolated from BEDT-TTF$^0$ to BEDT-TTF$^{+1}$ and the equation for $\nu_2(a_g)$ can be modified as follows:
\begin{equation}
\label{eq:modesplitting2}
\delta\nu_{2}  =   98 \frac{\rm cm^{-1}}{e} \delta\rho \quad .
\end{equation}

\begin{figure}[h]
    \centering
        \includegraphics[width=0.81\columnwidth]{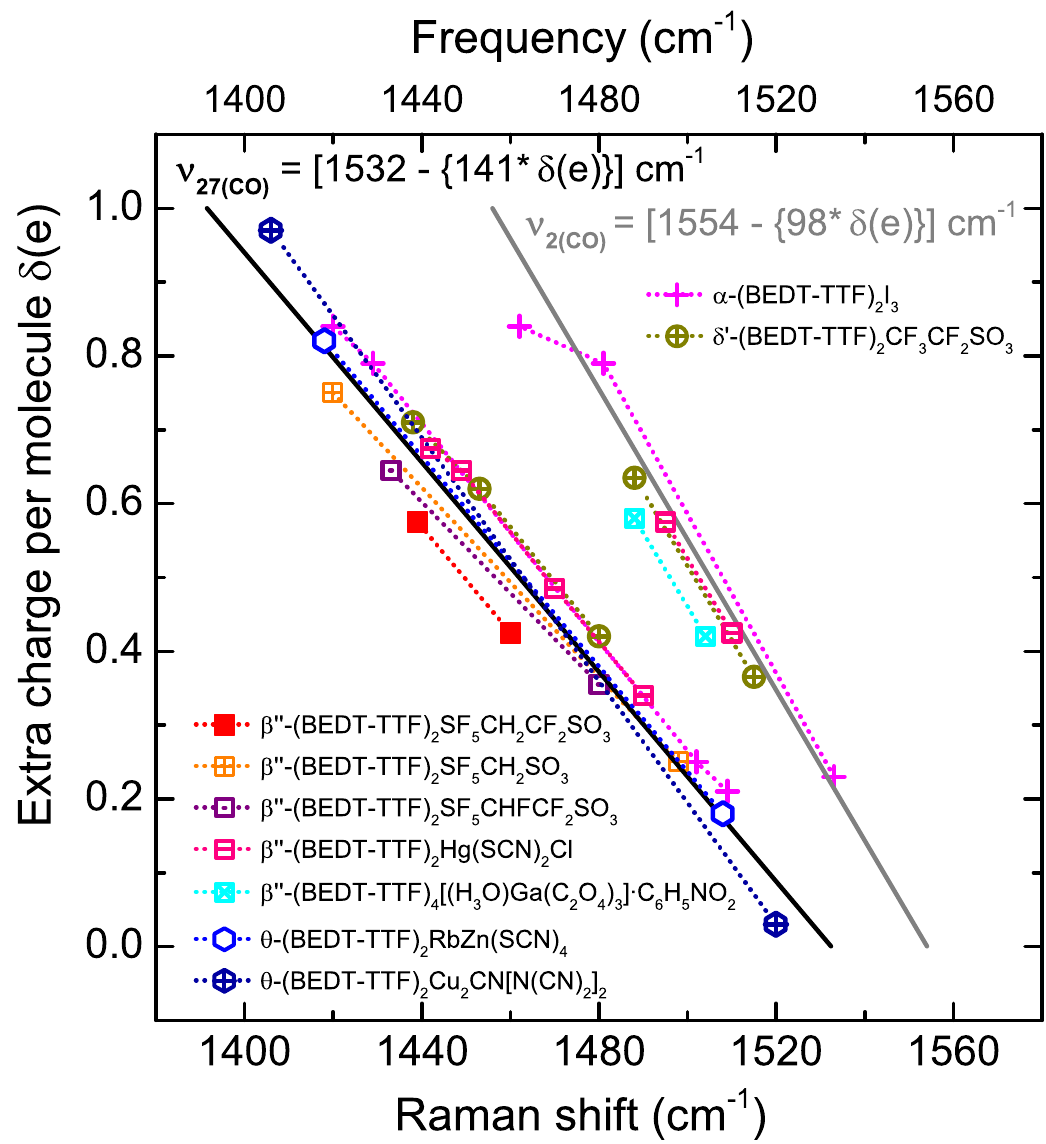}
        \caption{Positions of the $\nu_2(a_g)$ and the $\nu_{27}(b_{1u})$ modes in different BEDT-TTF-based salts for which charge disproportionation was identified by either Raman or infrared spectroscopy; or have asymmetric splitting (the solid lines are the relations obtained from \ref{fig:splitting1}). The exact compositions and frequencies are given in Table~\ref{tab:splitting}.}
    \label{fig:splitting}
\end{figure}

Fig.~\ref{fig:splitting} displays the spectral position of the $\nu_2$($a_g$) and the $\nu_{27}(b_{1u})$ modes in other BEDT-TTF-based salts for which charge disproportionation was identified by either Raman or infrared spectroscopy; or have multiple peaks (probably due to a more complex charge-ordered patterns or additional interactions in this state). The peak frequencies of the charge-ordered compounds, as plotted in Figs. \ref{fig:splitting1}, and \ref{fig:splitting} are listed in Table~\ref{tab:splitting}. Several $\theta$-phase compounds report strong coupling of the $\nu_2$($a_g$) to $\nu_3$($a_g$) modes in the charge-ordered state \cite{Yamamoto02, Suzuki04, Yamamoto05, Yamamoto04}, which are not plotted to reduce complexity. Applying $\delta\nu_{27}  =  140~\frac{\rm cm^{-1}}{e} \delta\rho$ \cite{Yakushi12}, and the consistent relation of $\nu_2$($a_g$) is: $\delta\nu_{2}  =  98~\frac{\rm cm^{-1}}{e} \delta\rho$, the charge disproportionation was determined for the compounds where the charge was not specified and plotted on the ordinate \footnote{Here we plot the vibrational modes observed in the charge-ordered state, which often is established only at low temperatures. This explains the sift among the different compounds.}. Assuming that the charges rearrange only among two molecules, in most cases the splitting is assumed to be symmetric to $\rho_0 = 0.5e$, it can be described as follows:
\begin{equation}
\delta\rho = \rho_{\rm rich} - \rho_{\rm poor} =
(\rho_0 + \tfrac{1}{2}\delta\rho) - (\rho_0 - \tfrac{1}{2}\delta\rho).
\label{eq:chargeimbalance}
\end{equation}

Often, more complex patterns are observed, leading to more than two peaks in charge ordered states. The most prominent charge-ordered BEDT-TTF system $\alpha$-(BEDT-TTF)$_3$I$_3$, for instance, records four distinct molecular charged sites \cite{Kakiuchi07,Ivek11,Tomic15}. The relative intensities of these peaks observed in infrared spectroscopy can be significantly different due to the dipole moments involved \cite{Girlando11,Girlando24}. For simplified discussion, we only consider the peak positions in charge-ordered compounds.

From the plotted values of charge disproportionation and the new slope for site charge, as illustrated in Fig.~\ref{fig:splitting}, it is crucial to note the frequency values for these two modes are spread over 20 \cm, which is originating presumably from the structural influences due to different stacking patterns, spectrometer resolution and temperature effects. For the comparison, we have selected the examples, which show prominent splitting with the onset of charge-order transition and not the examples which illustrate occurrence of a side bands or shoulder peaks at low temperatures. The relations in a symmetric charge-ordered state can be summarised as follows, which accounts for the splitting of the C=C stretching modes, where $\delta\rho$ is the charge imbalance:
\begin{subequations}
\label{eq:modesplitting}
\begin{eqnarray}
\delta\nu_{27} & = &  141 \frac{\rm cm^{-1}}{e} \delta\rho \quad , \\
\delta\nu_{2} & = &  98 \frac{\rm cm^{-1}}{e} \delta\rho \quad .
\end{eqnarray}
\end{subequations}

\section{Absolute charge}

\begin{figure}[h]
    \centering
        \includegraphics[width=\columnwidth]{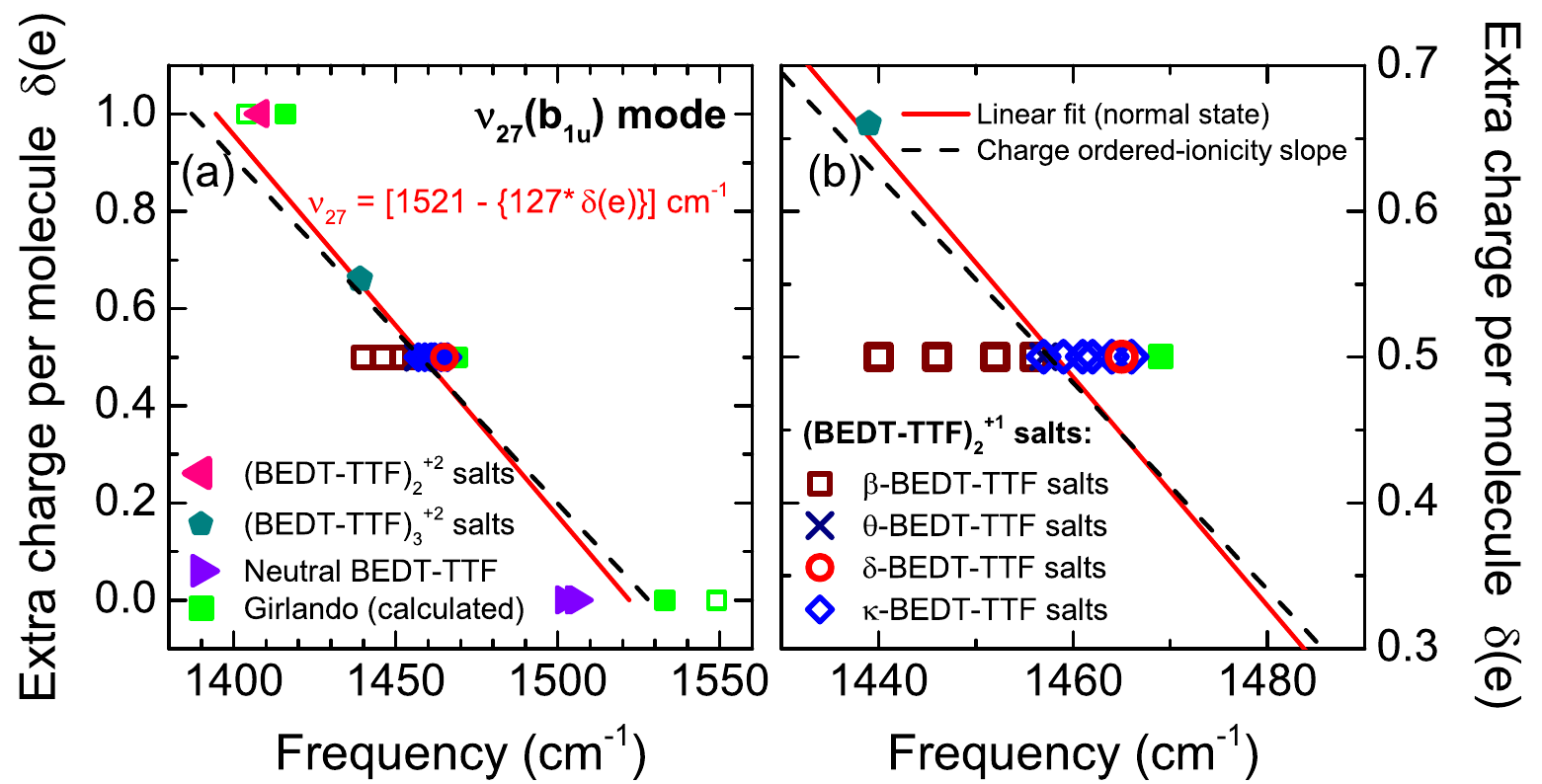}
        \caption{(a) Frequency of the $\nu_{27}$($b_{1u}$) mode determined from infrared measurements of different BEDT-TTF salts as indicated. For the respective charge per molecule the value is taken as assumed in the corresponding literature. The exact compositions and frequencies are listed in Table~\ref{tab:nu27}. Panel (b) is enlarged central part of panel (a). The red line is the linear fit for the experimental data and the black dashed line is the slope of charge-ordered state, shifted 4~\cm{} below to match central frequency for BEDT-TTF$^{+0.5}$. The green square symbols corresponds to the calculated values by Girlando \cite{Girlando11}, closed squares: dimer (BEDT-TTF)$_2$, and open symbols: monomer BEDT-TTF.}
    \label{fig:nu27}
\end{figure}

\begin{figure}
    \centering
        \includegraphics[width=\columnwidth]{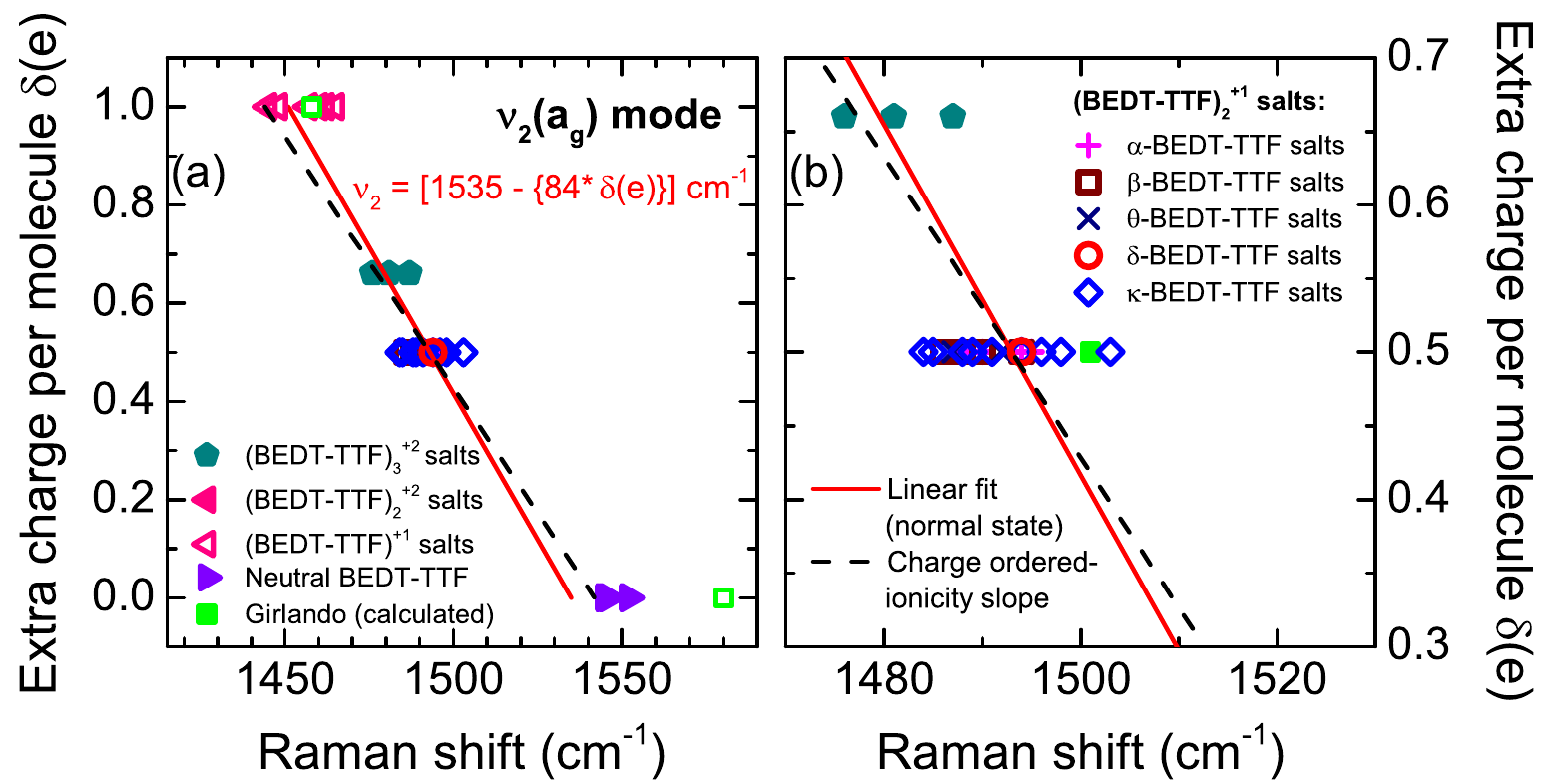}
        \caption{(a) Central frequency of the $\nu_2$($a_g$) mode determined from Raman measurements of different BEDT-TTF salts as indicated. For the respective charge per molecule the value is taken as assumed in the corresponding literature. The exact compositions and frequencies are listed in Table~\ref{tab:nu2}. Panel (b) is enlarged central part of panel (a). The red line is the linear fit for the experimental data and the black dashed line is the slope of charge-ordered state, shifted 12~\cm{} below to match central frequency for BEDT-TTF$^{+0.5}$. The strong deviations in the slopes and a large distribution for BEDT-TTF$^{+0.5}$ compounds will limit the detection of small deviations in charge. The green square symbols corresponds to the calculated values by Girlando \cite{Girlando11}, closed squares: dimer (BEDT-TTF)$_2$, and open symbols: monomer BEDT-TTF.}
    \label{fig:nu2}
\end{figure}

The situation becomes more complicated, when the {\em absolute value} of charge residing on the BEDT-TTF molecules becomes decisive. In order to determine the ionicity of BEDT-TTF from the known relations, it is imperative to consider the exact symmetry of BEDT-TTF molecules or dimers and technical experimental factors, including the resolution, shifts due to calibrations in Raman scattering measurements, calibrations, etc. Table~\ref{tab:nu27} summarizes the reported frequencies of the $\nu_{27}(b_{1u})$ mode observed by infrared spectroscopy in various BEDT-TTF salts. For a valid comparison, the numbers have been adjusted for temperature effects, by rescaling them according to Eq.~(\ref{eq:temperaturedependence}), details are given in Appendix~\ref{sec:TemperatureDependence}. Only the most intense peaks are considered for the plots when a side band or a shoulder peak is present \footnote{A side band in $\nu_{27}$($b_{1u}$) within a proximity of 5 \cm{} is not a signature of charge ordering, but arises due to unequivalent sites in a unit cell, as described in Ref. \cite{Sedlmeier12}}. The data for $\nu_{27}$($b_{1u}$) mode are displayed in Fig.~\ref{fig:nu27}, along with the vibrational frequencies of neutral and fully ionized molecules. Similar conclusions can be drawn when considering the Raman active mode, Fig.~\ref{fig:nu2} displays the findings on the $\nu_2$($a_g$) mode reported in the literature. As tabulated in Table~\ref{tab:nu2}, the investigations comprise different oxidation stages (according to stoichiometry) ($+0.5e$, and $+0.66e$), ranging from neutral molecules to fully ionized compounds ($+1e$).

\begin{figure}[h!]
    \centering
        \includegraphics[width=0.75\columnwidth]{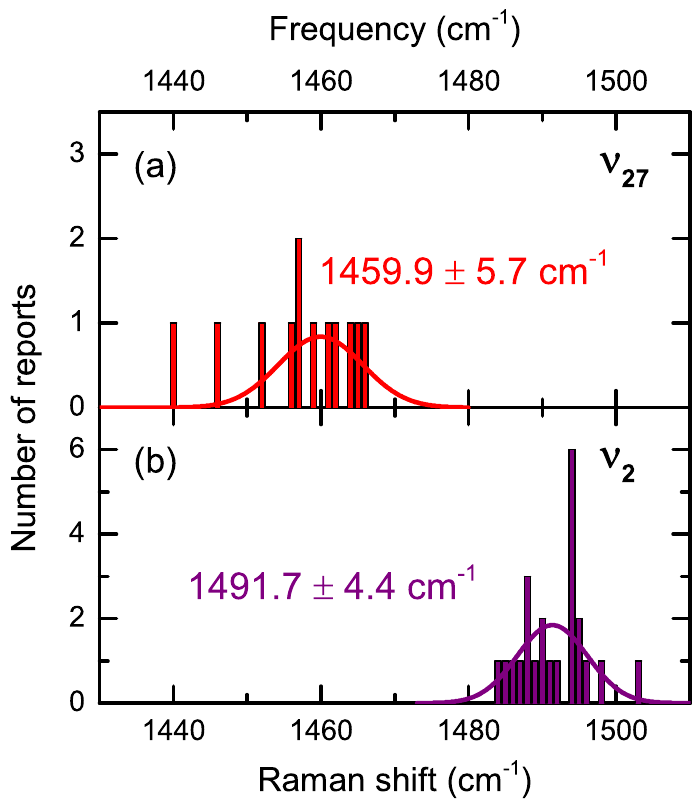}
        \caption{Histogram of the room-temperature frequencies reported for the (a) $\nu_{27}$($b_{1u}$) mode, and (b) $\nu_2$($a_g$) mode, obtained from infrared and Raman scattering experiments on different BEDT-TTF salts with nominally half an electron removed per molecule, BEDT-TTF$^{+0.5}$. If treated as a random deviation from the mean, the fit by a Gaussian distribution results in the center frequency and uncertainty as indicated.
        \label{fig:histogram}
        }
\end{figure}

When looking at the magnified scale on Fig.~\ref{fig:nu27}(b), Fig.~\ref{fig:nu2}(b), it becomes obvious that even for the nominally half oxidized BEDT-TTF, the findings spread is more than expected from the typical experimental accuracy. Since most of the compounds are characterized as partially ionized with half an electron removed per BEDT-TTF molecule, same as one electron donated by the two molecules of BEDT-TTF. If treated as a random deviation, it can be described by statistical probability for the $\nu_{27}$($b_{1u}$), and $\nu_2$($a_g$) modes in BEDT-TTF$^{+0.5}$ systems and we present our finding in Fig. \ref{fig:histogram}; compounds in charge ordered state are not considered for absolute charge determination. The frequencies found in the literature exhibit a considerable spread of 11.4~\cm, determined by the standard deviation assuming a Gaussian distribution around 1460~\cm, for $\nu_{27}$($b_{1u}$) and a spread of 8.8~\cm, with central frequency of 1491.7~\cm, for $\nu_2$($a_g$). No significant correlations can be seen between the vibrational frequencies and the crystallographic packing. The dimerized salts, in particular the $\kappa$-phase compounds, seem to be more homogeneous, {\it i.e.} the observed spectral distribution is narrower compared to the non-dimerized ones. For the BEDT-TTF salts with a 3:2 stoichiometry, (BEDT-TTF)$_3$Cl$_{2}\cdot$2H$_2$O  and (BEDT-TTF)$_3$(HSO$_4$)$_2$ \cite{Wang94,WilliamsBook}, the $\nu_2$ modes also observe a large spread of values. As already realized by Wang {\it et al.} \cite{Wang94,Wang96}, there seems to be no dependence of the frequency of the $\nu_2$($a_g$) vibrational mode on the crystallographic structure or molecular pattern. The salts with an non-stoichiometric anion layers are omitted, as exact charge from the linear regression might not be very accurate to determine slight changes accurately, given a distribution around 10 \cm, and considering the peak positions are very similar to 2:1 compounds.

The spread of central frequency modes for BEDT-TTF$^{+0.5}$, will equal to a charge distribution of about $0.089e$ for $\nu_{27}$($b_{1u}$) and $0.105e$ for $\nu_2$($a_g$). There may be many reasons for this: we cannot judge experimental errors notably the frequency calibration of the Raman shift and the resolution of the experiments. The stretching modes are known to couple to the electronic, molecular or structural environment leading to significant displacements of the vibrational frequencies \cite{Dressel04,Yamamoto05,Girlando11,Girlando24}. The molecular vibrations are affected by the electronic distribution commonly described by a coupling constant that might vary for the different compounds.  Moreover, the drastic change of slope between the charge-ordered state and normal state for the $\nu_2$($a_g$) mode, as illustrated in Fig. \ref{fig:nu2} is puzzling, and significantly alters the linear relations. Finally, the actual charge per molecule may be different for the various salts despite the fact that the oxidation state of $+0.5$ for the BEDT-TTF molecule was deduced from stoichiometric arguments. The effects of slight structural factors can be unambiguously realized with the reported difference of roughly 40 \cm{} in the calculated values for BEDT-TTF$^0$ molecule in $D_{2h}$ staggered conformation with the observed values for neutral eclipsed conformation of BEDT-TTF molecule \cite{Kozlov87, Girlando11} (the staggered and eclipsed conformation refers to the relative arrangement of the outer two rings of the BEDT-TTF molecule, as the inner rings are planar). Nevertheless, considering the large number of BEDT-TTF salts with different anions, the charge of $+1e$ per each two BEDT-TTF molecules seems to be fixed by the cations and slight shifts in the frequency of these modes in the normal state is not correlated with drastic changes in electronic properties. 

However, from other TTF-based compounds, it is known that the charge transfer might vary for salts formed by different anions \cite{YamadaBook}. Early studies of TTF (abbreviated for tetrathiafuran) salts report that these compounds can have non-integer charge transfer; in crystals with halides $X$ = Cl, Br, I, and other anions such as (SCN) or (SeCN), for instance, (TTF)$X_x$ appears with different stoichiometry $x = 0.57 - 0.79$ \cite{Warmack75,Somoano75,Sugano77,Wudl77,Tomkiewicz78,Kuzmany77,*Kuzmany79,Williams80,Madison82}. In particular Pecile and collaborators pioneered using frequencies of molecular vibrations for determining the degree of charge transfer in various radical ion systems \cite{Bozio79,Bozio80a,Bozio80b,Girlando83,Meneghetti84,Meneghetti86} in TTF based systems, which could in principle be extended to BEDT-TTF salts, but the different arrangements in BEDT-TTF based systems and an unusually large spread for stoichiometric BEDT-TTF$^{+0.5}$ salts, limits the confirmation of non-integer charge transfer in BEDT-TTF based systems.

\begin{table}
\caption{Charge sensitive C=C stretching modes.
From the Gaussian fit of the Fig. \ref{fig:histogram}, we derive the vibrational frequencies for BEDT-TTF$^{+0.5}$ and the uncertainty defined as full width at half maximum (FWHM). \label{tab:modes}
}
\begin{ruledtabular}
\begin{tabular}{l|ccc}
vibrational mode & $\nu_{27}(b_{1u})$&  $\nu_{2}(a_{g})$ \\
               & (cm$^{-1}$)       & (cm$^{-1}$)         \\
\hline
center frequency & 1459.9      &    1491.7           \\
confidence      & 11.4            & 8.8                  \\
relative precision & 0.78\%    & 0.58\%                  \\
\end{tabular}
\end{ruledtabular}
\end{table}

\section{$\nu_{3}$ mode}

\begin{figure}
    \centering
        \includegraphics[width=\columnwidth]{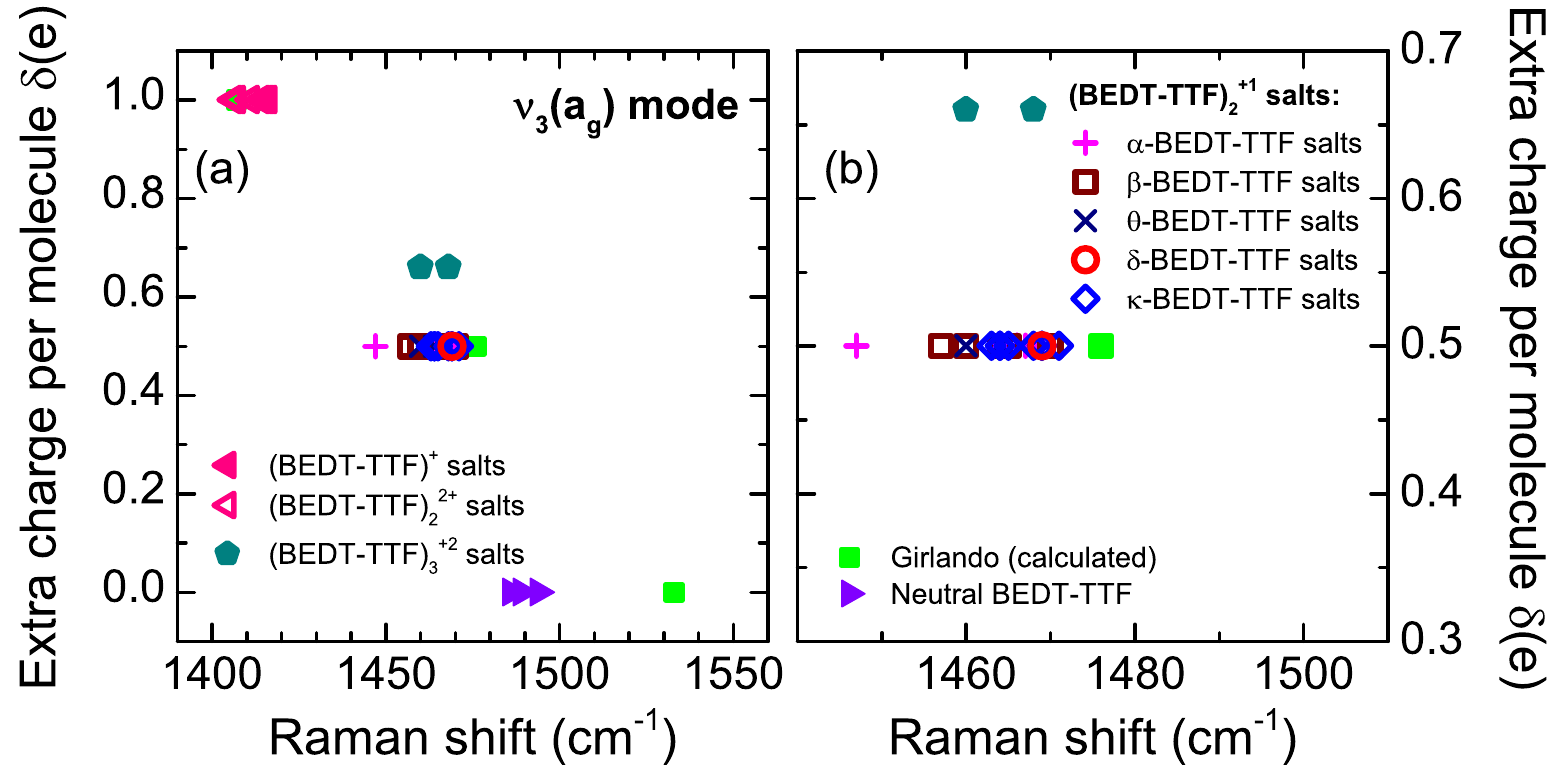}
        \caption{The frequency of the $\nu_3(a_g)$ stretching mode determined from Raman scattering experiments on different BEDT-TTF salts. For the respective charge per molecule the value is taken as assumed in the corresponding literature. The exact compositions and frequencies are listed in Table~\ref{tab:nu3}. Panel (b) is enlarged central part of panel (a).}
    \label{fig:nu3}
\end{figure}

\begin{figure}[h]
    \centering
        \includegraphics[width=0.7\columnwidth]{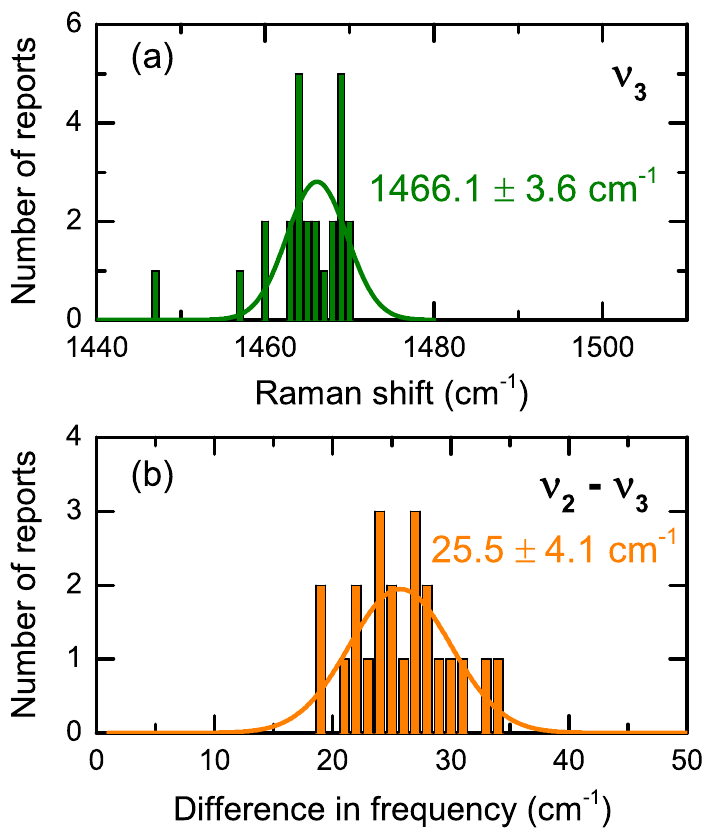}
        \caption{(a) Histogram of the room-temperature frequencies reported for the $\nu_3(a_g)$ Raman modes of different BEDT-TTF salts with nominally half an electron removed per molecule. Panel (b) displays the difference in frequency of the two Raman modes $\nu_2$($a_g$) and $\nu_3$($a_g$). The fit by a Gaussian distribution results in the center frequency and uncertainty as indicated in both panels. See Table~\ref{tab:nu3},\ref{tab:nu2-nu3} for the exact compositions and frequencies.
        \label{fig:nu3-diff}
        }
\end{figure}

Often, the second Raman mode confined to central C=C stretching vibration, the $\nu_3$($a_g$) mode, is considered to have a very similar dependence to the $\nu_2$($a_g$) mode by the theoretical studies \cite{Girlando11}, but the lack of any splitting of this mode in cases of charge-ordered states which record splitting of $\nu_2$($a_g$) and $\nu_{27}(b_{1u})$, indicate that $\nu_3$($a_g$) is not charge-sensitive. In fact, $\nu_3$ does strongly couples to the electronic response in dimer phases, producing shift of around 200 \cm{} for the infrared-active $\nu_3$ molecular vibrations in $b_{1u}$ and $b_{3u}$ geometry and a broad Raman mode in $b_{3g}$ geometry; only a single $\nu_3$ mode in $a_g$ symmetry is not coupled to the electronic response (see Ref. \cite{Maksimuk01}). Nevertheless, we plot the peak position values, and histogram for BEDT-TTF$^{+0.5}$ for $\nu_3$($a_g$) mode of different compounds in Figs.~\ref{fig:nu3}, and \ref{fig:nu3-diff}, the data is tabulated in Table~\ref{tab:nu3}. From our statistical analysis of about 25 different compounds reported in the literature, we obtain a vibrational frequency of $1464.6~(\pm 4.7)$~\cm, somewhat lower than the value given by Yamamoto {\it et al.} and Girlando \cite{Yamamoto05,Girlando11}. This comparison is for normal state values and does not include the specific examples of $\theta$-phase, which report strong coupling with $\nu_{2}$.

It is also interesting to consider only the difference between the to Raman modes, $\nu_{2}(a_{g})- \nu_{3}(a_{g})$, usually recorded in the same experiment and identical conditions; which removes the effects of calibrations in Raman measurements and is plotted for BEDT-TTF$^{+0.5}$ in Fig.~\ref{fig:nu3-diff}. The spread for the $\nu_{2}(a_{g})-\nu_{3}(a_{g})$ is also around 10 \cm, similar to that for $\nu_2(a_g)$, $\nu_{27}(b_{1u})$ and $\nu_3(a_g)$, which reduces the influence of calibrations in Raman measurements, and highlights a relatively high frequency distribution in experiments for different salts. It is also worth mentioning the strong emv-coupling to $\nu_3$ leads to lifting of degeneracy of $a_g$ and $b_{3g}$ modes, where $b_{1u}$ is observed in cross-polarisation in Raman spectra and the difference between the $a_g$ and $b_{3g}$ modes related to the strength of emv-coupling and is an assessment of electronic response and varies for different compounds. The identification of different modes in specific polarisation channels was done early on by extensive symmetry analysis, along with $^2$D and $^{13}$C isotope studies on several compounds, including \kCuBr, $\theta$-(BEDT-TTF)$_2$RbZn(SCN)$_4$, and $\theta$-(BEDT-TTF)$_2$Cu$_2$CN[N(CN)$_2$]$_2$ \cite{Maksimuk01, Yamamoto04, Suzuki05}.

In Table~\ref{tab:longtable} a complete list of all compounds is given for which vibrational frequencies of the C=C stretching modes were extracted by infrared or Raman experiments. The measurements were performed at different temperatures, as indicated. Since we are not in the position to judge technical issues and advances, specific experimental uncertainties, computational mistakes or other errors, we have confined ourselves in reproducing the values given in the respective literature.

\section{Summary and conclusions}

In summary, we have focused on the charge-sensitive vibrational modes in BEDT-TTF charge transfer salts. Considering the examples illustrating charge ordering where clear splitting of the charge-sensitive modes, i.e., infrared active $\nu_{27}(b_{1u})$ and Raman active $\nu_{2}(a_{g})$ mode are observed, following relations describe the the charge disproportionation, which are consistent with each other:
\begin{equation}
\delta\nu_{27}  =  141~\frac{\rm cm^{-1}}{e} \delta\rho \quad , \quad
\delta\nu_{2}   =  98~\frac{\rm cm^{-1}}{e} \delta\rho \quad .
\end{equation}

However, a significant spread in the frequencies of the vibrations complicates the application of these modes to calculate absolute values of charge per molecule, which highlights the limitation of this method to detect slight changes in the charge over the BEDT-TTF molecules, as would be the case for the non-stoichiometric examples, detection limit is at least around $\Delta\rho~\approx~(\pm~0.045)e$.

\begin{acknowledgments}
We appreciate the discussions with Kazushi Kanoda, Simone Fratini, Alberto Girlando, Takashi Yamamoto, and Maxim Wenzel. The project was supported by the Deutsche Forschungsgemeinschaft (DFG) via DR228/39-3 and DR228/74-1. The work at JHU was supported by NSF Award No. DMR-2004074.
\end{acknowledgments}

\section*{Data Availability} 
The data that support the findings of this study are available from the authors on reasonable request

\clearpage

\begin{table}
\centering
\caption{Splitting of the vibrational modes $\nu_2(a_g)$ and $\nu_{27}(b_{1u})$ in charge-ordered BEDT-TTF salts as determined from Raman and infrared spectroscopy. In some cases three of even four peak are identified, indicating a complex charge pattern \cite{Ward00,Schlueter01,Uruichi06,Ivek11,Yakushi12,Ivek17,Pustogow19}.
More details are given in Table~\ref{tab:longtable}.\label{tab:splitting}}
\begin{ruledtabular}
\begin{tabular}{l|cc}
\textbf{BEDT-TTF salts} & $\nu_2(a_{g})$       & $\nu_{27}(b_{1u})$  \\
                        & (cm$^{-1}$)   & (cm$^{-1}$) \\
\hline
$\alpha$-(BEDT-TTF)$_3$I$_3$  & 1462, 1481,~& 1420, 1429  \\
  &  1533 & 1502, 1509  \\
$\alpha^{\prime}$-(BEDT-TTF)$_2$IBr$_2$  & 1480, 1490,~&  1405, 1423 \\
  &  1545, 1556 & 1516  \\
$\beta^{\prime\prime}$-(BEDT-TTF)(TCNQ)  &  1479, 1535 & 1423, 1506 \\
$\beta^{\prime\prime}$-(BEDT-TTF)$_2$SF$_5$CH$_2$CF$_2$SO$_3$  &  & 1439, 1460 \\
$\beta^{\prime\prime}$-(BEDT-TTF)$_2$SF$_5$CH$_2$SO$_3$  &  & 1420, 1498 \\
$\beta^{\prime\prime}$-(BEDT-TTF)$_2$SF$_5$CHFCF$_2$SO$_3$  &   & 1433, 1480\\
$\beta^{\prime\prime}$-(BEDT-TTF)$_2$Hg(SCN)$_2$Cl & 	1495, 1510 & 1442, 1449,	\\
											& 				&  1470, 1490	\\
$\beta^{\prime\prime}$-(BEDT-TTF)$_4$ & 1488, 1504  &   \\
{           } [(H$_3$O)Ga(C$_2$O$_4$)$_3$]$\cdot$C$_6$H$_5$NO$_2$  &  &   \\
$\theta$-(BEDT-TTF)$_2$RbZn(SCN)$_4$  &    	& 	1418, 1508\\
$\theta$-(BEDT-TTF)$_2$Cu$_2$CN[N(CN)$_2$]$_2$ &     &  1407, 1520   \\
$\kappa$-(BEDT-TTF)$_2$Hg(SCN)$_2$Cl   &  1480, 1510  & 1440, 1442, \\
    &   & 1469, 1473 \\
$\delta'$-(BEDT-TTF)$_2$CF$_3$CF$_2$SO$_3$ &  1488, 1515  & 1438, 1453, \\
    &   & 1480 \\
\end{tabular}
\end{ruledtabular}
\end{table}

\begin{table}[h]
\centering
\caption{Vibrational frequency of the asymmetric $\nu_{27}(b_{1u})$ stretching mode of various BEDT-TTF compounds obtained from infrared spectroscopy. The data have been rescaled to $T=300$~K using Eq.~(\ref{eq:temperaturedependence}) in order to allow comparison. The data are plotted in Fig.~\ref{fig:nu27}; more details and references are given in Table~\ref{tab:longtable}.
\label{tab:nu27}}
\begin{ruledtabular}
\begin{tabular}{l|c}
\textbf{BEDT-TTF salts} & $\nu_{27}(b_{1u})$~~~(cm$^{-1}$)  \\
\hline
\textbf{(BEDT-TTF)$^{+0.5}$:}\\
$\beta^{\prime\prime}$-(BEDT-TTF)(TCNQ)  & 1452  \\
$\beta^{\prime\prime}$-(BEDT-TTF)$_2$AuBr$_2$  & 1456  \\
$\beta^{\prime\prime}$-(BEDT-TTF)$_2$SF$_5$CHFSO$_3$  &  1446 \\
$\beta^{\prime\prime}$-(BEDT-TTF)$_2$Hg(SCN)$_2$Cl & 	1440  \\
$\theta$-(BEDT-TTF)$_2$I$_3$  &   1457  \\
$\kappa$-(BEDT-TTF)$_2$Hg(SCN)$_2$Cl    &  1457  \\
$\kappa$-(BEDT-TTF)$_2$Hg(SCN)$_2$Br &   1459 \\
$\kappa$-(BEDT-TTF)$_2$Cu[N(CN)$_2$]Cl    &   1466  \\
$\kappa$-(BEDT-TTF)$_2$Cu[N(CN)$_2$]Br  & 1464  \\
$\kappa$-(BEDT-TTF)$_2$Cu$_2$(CN)$_3$  &  1461 \\
$\kappa$-(BEDT-TTF)$_2$Ag$_2$(CN)$_3$  &  1462 \\
$\delta^{\prime}$-(BEDT-TTF)$_2$CF$_3$CF$_2$SO$_3$ &  1465 \\
\hline
\textbf{(BEDT-TTF)$_2^{+2}$:}\\
(BEDT-TTF)$_2$[Mo$_6$O$_{19}$]  &  1454 \\
\hline
\textbf{(BEDT-TTF)$^{+0.66}$:}\\
$\beta^{\prime\prime}$(BEDT-TTF)$_3$(HSO$_4$)$_2$  & 1439  \\
\end{tabular}
\end{ruledtabular}
\end{table}

\begin{table}[h!]
\centering
\caption{Vibrational frequency of the $\nu_{2}(a_{g})$ stretching mode of various BEDT-TTF compounds obtained from Raman scattering experiments. The data have been rescaled to $T=300$~K using Eq.~(\ref{eq:temperaturedependence}) in order to allow comparison. Data data are visualized in Fig.~\ref{fig:nu2}. More details and references are given in Table~\ref{tab:longtable}.\label{tab:nu2}}
\begin{ruledtabular}
\begin{tabular}{l|c}
\textbf{BEDT-TTF salts} & $\nu_{2}(a_{g})$  \\
                         & (cm$^{-1}$) \\
\hline
\textbf{(BEDT-TTF)$^{+0.5}$:}\\
$\alpha$-(BEDT-TTF)$_4$[C(CN)$_2$CONH$_2$]CuCl$_2$  & 1495 \\
$\alpha$-(BEDT-TTF)$_4$[C(CN)$_2$CONH$_2$]CuCl$_2$  & 1494 \\
$\alpha$-(BEDT-TTF)$_4$[C(CN)$_2$CONH$_2$]CuBr$_2$  & 1494 \\
$\alpha$-(BEDT-TTF)$_4$[C(CN)$_2$CONH$_2$]CuBr$_2$  & 1495 \\
$\alpha$-(BEDT-TTF)$_2$(NH$_4$)Hg(SCN)$_4$  & 1488  \\
$\alpha$-(BEDT-TTF)$_2$RbHg(SCN)$_4$  &  1490 \\
$\beta$-(BEDT-TTF)$_2$I$_3$ & 1488 \\
$\beta$-(BEDT-TTF)$_2$I$_3$ & 1490 \\
$\beta^{\prime\prime}$-(BEDT-TTF)(TCNQ) (Metal)  & 1494 \\
$\beta^{\prime\prime}$-(BEDT-TTF)$_2$AuBr$_2$  &  1494 \\
$\beta^{\prime\prime}$-(BEDT-TTF)$_2$Hg(SCN)$_2$Cl & 	1494	\\
$\beta^{\prime\prime}$-(BEDT-TTF)$_4$[(H$_3$O)Ga(C$_2$O$_4$)$_3$]$\cdot$C$_6$H$_5$NO$_2$  &  1486 \\
$\theta$-(BEDT-TTF)$_2$CsZn(SCN)$_4$    &  1487  \\
$\theta$-(BEDT-TTF)$_2$I$_3$  &  1492 \\
$\kappa$-(BEDT-TTF)$_2$Hg(SCN)$_2$Cl &  1485  \\
$\kappa$-(BEDT-TTF)$_2$Hg(SCN)$_2$Br &  1484   \\
$\kappa$-(BEDT-TTF)$_2$Cu[N(CN)$_2$]Cl &  1489                          \\
$\kappa$-(BEDT-TTF)$_2$Cu[N(CN)$_2$]Br &  1488                           \\
$\kappa$-(BEDT-TTF)$_2$Cu[N(CN)$_2$]Br &   1491  \\
$\kappa$-(BEDT-TTF)$_2$Cu[N(CN)$_2$]Br &  1496 \\
$\kappa$-(BEDT-TTF)$_2$Cu(NCS)$_2$  & 1503 \\
$\kappa$-(BEDT-TTF)$_2$Cu$_2$(CN)$_3$  &  1498 \\
$\delta^{\prime}$-(BEDT-TTF)$_2$CF$_3$CF$_2$SO$_3$ & 1494 \\
\hline
\textbf{(BEDT-TTF)$_2^{+2}$:}\\
(BEDT-TTF)$_2$[Mo$_6$O$_{19}$]  & 1460 \\
\hline
\textbf{(BEDT-TTF)$^{+}$:}\\
(BEDT-TTF)Cu[N(CN)$_2$]$_2$ & 1448 \\
(BEDT-TTF)BiI$_4$  & 1465 \\
(BEDT-TTF)AuBr$_2$Cl$_2$ & 1457  \\
(BEDT-TTF)AuBr$_2$Cl$_2$ & 1462 \\
(BEDT-TTF)AuBr$_2$Cl$_2$ &  1445 \\
(BEDT-TTF)(ClO$_4$)  & 1445 \\
\hline
\textbf{(BEDT-TTF)$^{+0.66}$:}\\
(BEDT-TTF)$_3$Cl$_2$.2H$_2$O    & 1487 \\
(BEDT-TTF)$_3$(HSO$_4$)$_2$  & 1476  \\
$\beta^{\prime\prime}$-(BEDT-TTF)$_3$(HSO$_4$)$_2$  & 1481  \\
\end{tabular}
\end{ruledtabular}
\end{table}

\begin{table}[h!]
\centering
\caption{Center frequency of the $\nu_{3}(a_{g})$ modes plotted in Fig.~\ref{fig:nu3} for various BEDT-TTF salts.
For comparison the values are rescaled to $T=300$~K
according to Eq.~(\ref{eq:temperaturedependence}).
More details and references are given in Table~\ref{tab:longtable}.\label{tab:nu3}}
\begin{ruledtabular}
\begin{tabular}{l|c}
\textbf{BEDT-TTF salts} & $\nu_{3}(a_{g})$  \\
                         & (cm$^{-1}$) \\
\hline
\textbf{(BEDT-TTF)$^{+0.5}$:}\\
$\alpha$-(BEDT-TTF)$_2$I$_3$ (Metal)  & 1464 \\
$\alpha$-(BEDT-TTF)$_4$[C(CN)$_2$CONH$_2$]CuCl$_2$  & 1466 \\
$\alpha$-(BEDT-TTF)$_4$[C(CN)$_2$CONH$_2$]CuCl$_2$  & 1467 \\
$\alpha$-(BEDT-TTF)$_4$[C(CN)$_2$CONH$_2$]CuBr$_2$  & 1466 \\
$\alpha$-(BEDT-TTF)$_4$[C(CN)$_2$CONH$_2$]CuBr$_2$  & 1468 \\
$\alpha$-(BEDT-TTF)$_2$(NH$_4$)Hg(SCN)$_4$  & 1469 \\
$\alpha$-(BEDT-TTF)$_2$RbHg(SCN)$_4$  & 1469\\
$\alpha^{\prime}$-(BEDT-TTF)$_2$IBr$_2$ (Metal)& 1447 \\
$\beta$-(BEDT-TTF)$_2$I$_3$ & 1457 \\
$\beta$-(BEDT-TTF)$_2$I$_3$ & 1465 \\
$\beta$-(BEDT-TTF)$_2$AuI$_2$ & 1470\\
$\beta^{\prime\prime}$-(BEDT-TTF)(TCNQ) (Metal)  &  1460 \\
$\beta^{\prime\prime}$-(BEDT-TTF)$_2$Hg(SCN)$_2$Cl & 	1470	\\
$\beta^{\prime\prime}$-(BEDT-TTF)$_4$[(H$_3$O)Ga(C$_2$O$_4$)$_3$]$\cdot$ C$_6$H$_5$NO$_2$  & 1464 \\
$\theta$-(BEDT-TTF)$_2$CsZn(SCN)$_4$ & 1464 \\
$\theta$-(BEDT-TTF)$_2$I$_3$  &  1460 \\
$\kappa$-(BEDT-TTF)$_2$Hg(SCN)$_2$Cl  & 1464 \\
$\kappa$-(BEDT-TTF)$_2$Hg(SCN)$_2$Br &  1465 \\
$\kappa$-(BEDT-TTF)$_2$Cu[N(CN)$_2$]Cl    & 1463 \\
$\kappa$-(BEDT-TTF)$_2$Cu[N(CN)$_2$]Br  &  1464 \\
$\kappa$-(BEDT-TTF)$_2$Cu[N(CN)$_2$]Br    & 1469 \\
$\kappa$-(BEDT-TTF)$_2$Cu[N(CN)$_2$]Br  &  1468 \\
$\kappa$-(BEDT-TTF)$_2$Cu(NCS)$_2$  & 1469 \\
$\kappa$-(BEDT-TTF)$_2$Cu$_2$(CN)$_3$  & 1471 \\
$\delta^{\prime}$-(BEDT-TTF)$_2$CF$_3$CF$_2$SO$_3$&  1469 \\
\hline
\textbf{(BEDT-TTF)$_2^{+2}$:}\\
(BEDT-TTF)$_2$[Mo$_6$O$_{19}$]  & 1414 \\
\hline
\textbf{(BEDT-TTF)$^{+}$:}\\
(BEDT-TTF)Cu[N(CN)$_2$]$_2$ & 1411 \\
(BEDT-TTF)BiI$_4$  & 1407 \\
(BEDT-TTF)AuBr$_2$Cl$_2$ & 1406 \\
(BEDT-TTF)AuBr$_2$Cl$_2$ & 1416 \\
(BEDT-TTF)AuBr$_2$Cl$_2$ & 1415 \\
(BEDT-TTF)(ClO$_4$)  &  1415 \\
\hline
\textbf{(BEDT-TTF)$^{+0.66}$:}\\
(BEDT-TTF)$_3$Cl$_2$.2H$_2$O    &  1468 \\
(BEDT-TTF)$_3$(HSO$_4$)$_2$  & 1460 \\
\end{tabular}
\end{ruledtabular}
\end{table}

\begin{table}[h]
\centering
\caption{Difference of the frequencies reported for $\nu_{2}(a_{g})$ and $\nu_{3}(a_{g})$ modes  for various BEDT-TTF salts.
More details and references are given in Table~\ref{tab:longtable}.\label{tab:nu2-nu3}}
\begin{ruledtabular}
\begin{tabular}{l|c}
\textbf{BEDT-TTF salts} & $\nu_{2}(a_{g})- \nu_{3}(a_{g})$  \\
                         & (cm$^{-1}$) \\
\hline
\textbf{(BEDT-TTF)$_2^{+}$:}\\
$\alpha$-(BEDT-TTF)$_4$[C(CN)$_2$CONH$_2$]CuCl$_2$  & 29 \\
$\alpha$-(BEDT-TTF)$_4$[C(CN)$_2$CONH$_2$]CuCl$_2$  & 27 \\
$\alpha$-(BEDT-TTF)$_4$[C(CN)$_2$CONH$_2$]CuBr$_2$  & 28 \\
$\alpha$-(BEDT-TTF)$_4$[C(CN)$_2$CONH$_2$]CuBr$_2$  & 27 \\
$\alpha$-(BEDT-TTF)$_2$(NH$_4$)Hg(SCN)$_4$  & 19 \\
$\alpha$-(BEDT-TTF)$_2$RbHg(SCN)$_4$  & 21 \\
$\beta$-(BEDT-TTF)$_2$I$_3$ & 31 \\
$\beta$-(BEDT-TTF)$_2$I$_3$ & 25 \\
$\beta^{\prime\prime}$-(BEDT-TTF)(TCNQ) (Metal)  &  33 \\
$\beta^{\prime\prime}$-(BEDT-TTF)$_2$Hg(SCN)$_2$Cl & 		24	\\
$\beta^{\prime\prime}$-(BEDT-TTF)$_4$[(H$_3$O)Ga(C$_2$O$_4$)$_3$]$\cdot$ & 22 \\
{~~~~} $\cdot$C$_6$H$_5$NO$_2$  &  \\
$\theta$-(BEDT-TTF)$_2$CsZn(SCN)$_4$ & 24 \\
$\theta$-(BEDT-TTF)$_2$I$_3$  &  30 \\
$\kappa$-(BEDT-TTF)$_2$Hg(SCN)$_2$Cl  & 22 \\
$\kappa$-(BEDT-TTF)$_2$Hg(SCN)$_2$Br &  19 \\
$\kappa$-(BEDT-TTF)$_2$Cu[N(CN)$_2$]Cl    & 26 \\
$\kappa$-(BEDT-TTF)$_2$Cu[N(CN)$_2$]Br  &  24 \\
$\kappa$-(BEDT-TTF)$_2$Cu[N(CN)$_2$]Br    & 23 \\
$\kappa$-(BEDT-TTF)$_2$Cu[N(CN)$_2$]Br  &  28 \\
$\kappa$-(BEDT-TTF)$_2$Cu(NCS)$_2$  & 34 \\
$\kappa$-(BEDT-TTF)$_2$Cu$_2$(CN)$_3$  & 27 \\
$\delta^{\prime}$-(BEDT-TTF)$_2$CF$_3$CF$_2$SO$_3$&  25 \\
\hline
\textbf{(BEDT-TTF)$^{+x}$:}\\
(BEDT-TTF)Hg$_{0.776}$(SCN)$_2$ & 20 \\
(BEDT-TTF)Ag$_{1.6}$(SCN)$_2$ &  24 \\
$\kappa$-(BEDT-TTF)$_4$Hg$_{3-\delta}$Cl$_8$  & 22 \\
\end{tabular}
\end{ruledtabular}
\end{table}
\clearpage

\onecolumngrid
\begin{longtable}{l|ccccc}
\caption{Molecular vibrations of C=C symmetric and asymmetric stretching modes for different BEDT-TTF based charge-transfer salts. Where important, experimental details are provided (measurements on crystals and powder; charge ordered or metallic, theoretical calculations; scattering configuration; here  ET stands for BEDT-TTF and BETS stands for BEDT-TSF). For each set of frequencies, the temperature in Kelvin  is added in parenthesis, RT refers to room temperature.
\label{tab:longtable}} \\
\hline\hline
\textbf{BEDT-TTF salts} & Space & $\nu_{27}$ (b$_{1u}$) (cm$^{-1}$) & $\nu_{2}$ (a$_{g}$) (cm$^{-1}$) & $\nu_{3}$ (a$_{g}$) (cm$^{-1}$) & Ref. \\
 &  group	& infrared active mode &    \multicolumn{2}{c}{Raman active modes}  &    \\
\hline

\endfirsthead
\caption{Molecular vibrations of C=C symmetric and asymmetric stretching modes for different BEDT-TTF based charge-transfer salts. (continued) \label{tab:longtable}} \\
\hline\hline
\textbf{BEDT-TTF salts} & Space & $\nu_{27}$ (b$_{1u}$) (cm$^{-1}$) & $\nu_{2}$ (a$_{g}$) (cm$^{-1}$) & $\nu_{3}$ (a$_{g}$) (cm$^{-1}$) & Ref. \\
 &  group	& infrared active mode &    \multicolumn{2}{c}{Raman active modes}  &   \\
\hline
\endhead
Neutral ET (crystal)& \textit{P}2$_1$/c & 1505 (RT) & 1552 (RT) & 1494 (RT) & \cite{WilliamsBook,Kozlov87} \\
Neutral ET-d$_8$ (crystal)& - & 1506 (RT) & 1552 (RT) & 1494 (RT) & \cite{Kozlov87} \\
Neutral ET (powder)& - & 1508 (5)  & - & - & \cite{Moldenhauer93b} \\
Neutral ET  & - & - & 1546 (RT) & 1489 (RT) & \cite{Wang94}  \\
Neutral ET (powder)  & - & - & 1554 (RT) & 1494 (RT) & \cite{Meneghetti86}  \\
ET$^0$ (calc.) & - & 1549 & 1580 & 1533  & \cite{Girlando11}  \\
(ET)$_2^0$ (calc.) & - & 1533 & 1565 & 1533  & \cite{Girlando11}  \\
ET$_2^+$ (calc.) & - &  1469 & 1501 & 1476& \cite{Girlando11}  \\
ET$_2^{2+}$ (calc.) & - & 1413 & 1457 &  1415 & \cite{Girlando11}  \\
ET$^+$ (calc.) & - & 1404 & 1458 & 1407 & \cite{Girlando11}  \\
\hline
\textbf{$\alpha$-phase (2:1 salts)}\\
$\alpha$-(ET)$_2$I$_3$ (metal) & \textit{P}$\overline{1}$ & 1448, 1460, 1479 (300) & 1487, 1511 (150) & 1470 (150) & \citep{Ivek11, Wojciechowski03} \\
$\alpha$-(ET)$_2$I$_3$ (CO) & \textit{P}1 & 1420, 1429,  & 1462, 1481, 1533 (20) & 1476, 1458 (20) &  \citep{Ivek11, Wojciechowski03} \\
  &   &  1502, 1509 (50)   &   & &   \\
$\alpha$-(ET)$_4$[C(CN)$_2$CONH$_2$]CuCl$_2$ & \textit{I}222 & - & 1495 (0\textdegree) (RT) & 1466 (0\textdegree) (RT) & \cite{Wang96s} \\
  &   & - & 1494 (90\textdegree) (RT) & 1467 (90\textdegree) (RT)  & \cite{Wang96s} \\
$\alpha$-(ET)$_4$[C(CN)$_2$CONH$_2$]CuBr$_2$ & \textit{I}222 & - & 1494 (0\textdegree) (RT) & 1466 (0\textdegree) (RT) & \cite{Wang96s} \\
   &   & - & 1495 (90\textdegree) (RT)& 1468 (90\textdegree) (RT) & \cite{Wang96s} \\
$\alpha$-(ET)$_2$(NH$_4$)Hg(SCN)$_4$ & \textit{P}$\overline{1}$ & - & 1488 (RT) & 1469 (RT) & \cite{WilliamsBook, Wang94} \\
$\alpha$-(ET)$_2$RbHg(SCN)$_4$ & \textit{P}$\overline{1}$ &  - & 1490 (RT) & 1469 (RT) & \cite{Wang94} \\
$\alpha^{\prime}$-(ET)$_2$IBr$_2$ (metal) & \textit{P}$\overline{1}$ &  1404, 1510 (250) & 1489, 1545, 1551 (250) & 1450 (250) & \cite{WilliamsBook,Yue08,Yue09} \\
$\alpha^{\prime}$-(ET)$_2$IBr$_2$ (CO) &   &  1405, 1423, 1516 (6) & 1480, 1490, & 1406, 1451, 1468 (20) & \cite{Yue08,Yue09} \\
  &  &   & 1545, 1556 (20) &   &   \\
\hline
\textbf{$\beta$-phase (2:1 salts)}\\
$\beta$-(ET)$_2$I$_3$ & \textit{P}$\overline{1}$ & - & 1488 (RT) & 1457 (RT) & \cite{Ferraro87,Wang94}\\
  &   & - & 1490 (RT) & 1465 (RT) & \cite{Eldridge02} \\
$\beta$-(ET)$_2$I$_3$-\textit{d}$_8$ &   & - & 1494 (RT) & 1467 (RT) & \cite{Eldridge02} \\
$\beta$-(ET)$_2$AuI$_2$ & \textit{P}$\overline{1}$ & - & - & 1470 (RT) & \cite{WilliamsBook, Wang94} \\
$\beta^{\prime\prime}$-(ET)(TCNQ) (metal) & \textit{P}$\overline{1}$ & 1460 (6) & 1501 (10) & 1468 (10) & \cite{Yakushi12, Uruichi06} \\
$\beta^{\prime\prime}$-(ET)(TCNQ) (CO) &   & 1423, 1506 (280) & 1479, 1535 (280) & 1330, 1457 (280) & \cite{Yakushi12, Uruichi06} \\
$\beta^{\prime\prime}$-(ET)$_2$AuBr$_2$ &   &  1464 (10)  & 1497 (10) & -& \cite{Yamamoto05} \\
$\beta^{\prime\prime}$-(ET)$_2$SF$_5$CH$_2$CF$_2$SO$_3$ (CO) & \textit{P}$\overline{1}$ & 1439, 1460 (300) & - & - & \cite{Ward00, Pustogow19} \\
$\beta^{\prime\prime}$-(ET)$_2$SF$_5$CH$_2$SO$_3$ (CO) & \textit{P}$\overline{1}$ &  1420, 1498 (300)  & - & -& \cite{Ward00, Pustogow19} \\
$\beta^{\prime\prime}$-(ET)$_2$SF$_5$CHFCF$_2$SO$_3$ (CO) & \textit{P}$\overline{1}$ & 1433, 1480 (300) & - & - & \cite{Schlueter01, Pustogow19} \\
$\beta^{\prime\prime}$-(ET)$_2$SF$_5$CHFSO$_3$ & \textit{P}$\overline{1}$ & 1452 (100) & - & - & \cite{Ward00, Pustogow19} \\
$\beta^{\prime\prime}$-(ET)$_2$Hg(SCN)$_2$Cl & \textit{P}$\overline{1}$ & 1440 (300) & 1500 (100) & 1470 (300) & \cite{li2017metal} \\
$\beta^{\prime\prime}$-(ET)$_2$Hg(SCN)$_2$Cl (CO) &  & 1442, 1449, & 1495, 1510 (5) & 1475 (5) & \cite{li2017metal} \\
												 &  & 1470, 1490 (10) & - & - &  \\
$\beta^{\prime\prime}$-(ET)$_4$[(H$_3$O)$M$(C$_2$O$_4$)$_3$]$\cdot Y$& \textit{C}2/\textit{c} & - & 1490 (200) & 1468 (200) & \cite{Akutsu02, Bangura05} \\
$\beta^{\prime\prime}$-(ET)$_4$[(H$_3$O)$M$(C$_2$O$_4$)$_3$]$\cdot Y$ (CO) &   & - & 1488, 1504 (13) & 1471 (13) & \cite{Bangura05} \\
${~~~~}$	($M$ = Ga, $Y$ = C$_6$H$_5$NO$_2$) &  &  &  &  &  \\
\hline
\textbf{$\theta$-phase (2:1 salts)}\\
$\theta$-(ET)$_2$RbZn(SCN)$_4$ (CO)$^*$ & \textit{P}2$_{1}$2$_{1}$2$_{1}$ & 1418, 1508 (10) & 1480, 1543 (10) & -  & \cite{Yamamoto05} \\
$\theta$-(ET)$_2$RbZn(SCN)$_4$ (CO)$^*$ &  & - & 1480, 1539, 1548 (50) & 1455+multiple (50)  & \cite{Yamamoto02} \\
$\theta_o$-(ET)$_2$TlZn(SCN)$_4$ (CO)$^*$ &   & - & 1479, 1540, 1550 (50) & 1454 (+multiple) (50)  & \cite{Suzuki04} \\
$\theta_o$-(ET)$_2$TlZn(SCN)$_4$ (metal) &   & - & 1486 (240) & -  & \cite{Suzuki04} \\
$\theta_m$-(ET)$_2$TlZn(SCN)$_4$ (CO)$^*$ &   & - & 1479, 1542, 1550 (50) & 1456 (+multiple) (50)  & \cite{Suzuki04} \\
$\theta_m$-(ET)$_2$TlZn(SCN)$_4$ (metal) &   & - & 1488, 1548 (220) & 1460 (200)  & \cite{Suzuki04} \\
$\theta$-(ET)$_2$CsZn(SCN)$_4$ & \textit{I}222 & 1458 (6) & 1495 (20) & 1471 (20) & \cite{Suzuki05} \\
$\theta$-(ET)$_2$I$_3$ & \textit{P}2$_1$/c & 1462 (150)  & 1492 (300) & 1460 (300) & \cite{Yakushi12,Yakushi06, Pustogow17} \\
$\theta$-(ET)$_2$Cu$_2$CN[N(CN)$_2$]$_2$ (CO)$^*$ & \textit{P}2$_1$/\textit{c}  & 1406, 1520 (100)  & 1490, 1550 (100) & 1400, 1455 (100) & \cite{Yamamoto05} \\
$\theta$-(ET)$_2$Cu$_2$CN[N(CN)$_2$]$_2$ (CO)$^*$ & -  & 1407, 1415, 1520 (100)  & 1489, 1545, 1555 (50) & 1450(+multiple) (50) & \cite{Yamamoto04} \\
$\theta$-(ET)$_2$Cu$_2$CN[N(CN)$_2$]$_2$ & -  & -  & 1470, 1540 (300) & 1460 (300) & \cite{Yamamoto04} \\
\hline

\pagebreak

\textbf{$\kappa$-phase (2:1 salts)}\\
$\kappa$-(ET)$_2$Hg(SCN)$_2$Cl & \textit{C}2/\textit{c} &  1457, 1462 (300) & 1492 (50) & 1470 (100) & \cite{Hassan18, Ivek17, Liebman25}\\
$\kappa$-(ET)$_2$Hg(SCN)$_2$Cl (CO) &   & 1440, 1442,  &  1480, 1510 (20) & 1470 (20)& \cite{Ivek17, Hassan18, Liebman25}\\
  &   & 1469, 1473 (10)  &  &  &  \\
$\kappa$-(ET)$_2$Hg(SCN)$_2$Br & \textit{C}2/\textit{c} & 1459 (300) & 1484 (300) & 1465 (300) & \cite{Hassan18, Ivek17}\\
$\kappa$-(ET)$_2$Cu[N(CN)$_2$]Cl & \textit{Pnma} & 1465 (300) & 1489 (RT) & 1463 (RT) & \cite{WilliamsBook, Wang94, Pinteric18}\\
$\kappa$-(ET)$_2$Cu[N(CN)$_2$]Br-\textit{d}$_8$ &   & & 1494 (RT) &  1470 (RT)  & \cite{Eldridge02}\\
$\kappa$-(ET)$_2$Cu[N(CN)$_2$]Br  & \textit{Pnma} & 1464 (300) & 1488 (RT) & 1464 (RT) & \cite{WilliamsBook, Wang94, Sedlmeier12}\\
								  &   &  & 1499 (15) & 1476 (15)  & \cite{Maksimuk01}\\
								  &   &  & 1496 (RT) & 1468 (RT)  & \cite{Eldridge02}\\
$\kappa$-(ET)$_2$Cu[N(CN)$_2$]I & \textit{Pnma} & - & 1503 (multiple) (5) & 1475 (multiple) (5) & \cite{Yamamoto24}\\
$\kappa$-(ET)$_2$Cu(NCS)$_2$ & \textit{P}2$_1$ & - & 1503 (RT) & 1469 (RT) & \cite{WilliamsBook, Eldridge02}\\
$\kappa$-(ET)$_2$Cu(NCS)$_2$-\textit{d}$_8$ &   & - & 1504 (RT) & 1469 (RT) & \cite{Eldridge02}\\
$\kappa$-(ET)$_2$Cu$_2$(CN)$_3$ & \textit{P}2$_1$/\textit{c} & 1461 (300) & 1498 (RT) & 1471 (RT)  & \cite{Geiser91b, Wang94,Sedlmeier12}\\
$\kappa$-(ET)$_2$Ag$_2$(CN)$_3$ & \textit{P}2$_1$/\textit{c} &  1457, 1462 (300) & - & - & \cite{Hiramatsu17, Pinteric18}\\
$\kappa$-(ET)$_2$[BR/S(salicylate)$_2$] & \textit{P}2$_1$ &  - & 1502, 1508 (5) & 1460, 1470 (5)& \cite{Blundell25}\\
\hline
\textbf{Non-stoichiometric}\\
(ET)Hg$_{0.776}$(SCN)$_2$ & \textit{P}$\overline{1}$ & - & 1500 (RT) &  1480 (RT) & \cite{Coppens91, Wang94} \\
(ET)Ag$_{1.6}$(SCN)$_2$ & \textit{Pbca} & -  & 1500 (RT) & 1476 (RT) & \cite{WilliamsBook, Wang94} \\
$\kappa$-(ET)$_4$Hg$_{2.89}$Br$_8$ & \textit{C}2/\textit{m} & 1454 (300) & 1494 (300) & - & \cite{Li98, Wang94}\\
$\kappa$-(ET)$_4$Hg$_{3-\delta}$Cl$_8$ &  \textit{I}2/\textit{c} &  - & 1489 (RT) & 1462 (RT) & \cite{Wang94, WilliamsBook}\\
				 &  					 &  - & 1486 (RT) & 1464 (RT) & \cite{Wang94}\\
\hline
\textbf{Other 2:1 ET systems}\\
$\delta'$-(ET)$_2$CF$_3$CF$_2$SO$_3$ & \textit{P}2$_1$/\textit{m} &  1465 (300) & 1494 (300) & 1469 (300) & \cite{Olejniczak22} \\
$\delta'$-(ET)$_2$CF$_3$CF$_2$SO$_3$ (CO) & - &  1438, 1453, 1480 (10) & 1488, 1515 (80) & 1474 (80) & \cite{Olejniczak22} \\
\hline
 \textbf{3:2 salts}\\
(ET)$_3$Cl$_2$.2H$_2$O & \textit{P}$\overline{1}$ &  - & 1487 (RT) & 1468 (RT)& \cite{WilliamsBook, Wang94} \\
(ET)$_3$(HSO$_4$)$_2$ & \textit{P}$\overline{1}$ & -  & 1476 (RT) & 1460 (RT) & \cite{WilliamsBook, Wang94}\\
$\beta^{\prime\prime}$-(ET)$_3$(HSO$_4$)$_2$ (Metal) &   &  1443 (200)  & 1485 (200)& - & \cite{Yamamoto05}\\
$\beta^{\prime\prime}$-(ET)$_3$(HSO$_4$)$_2$ (CO) &   &  1422, 1490 (100)  & 1474, 1527 (100)& - & \cite{Yamamoto05}\\
\hline
\textbf{1:1 salts}\\
(ET)$_2$[Mo$_6$O$_{19}$]  \{(ET)$_2^{+2}$\} & \textit{P}2$_1$/\textit{c} & 1408 (RT) & 1460 (RT) & 1414 (RT) & \cite{Visentini98} \\
(ET)Cu[N(CN)$_2$]$_2$ & - &  - & 1448 (RT) & 1411 (RT) & \cite{Wang94}\\
(ET)BiI$_4$ & \textit{P}$\overline{1}$ & -  & 1465 (RT) & 1407 (RT) & \cite{WilliamsBook, Wang94} \\
(ET)AuBr$_2$Cl$_2$ & \textit{Pnnm} & - & 1457 (RT) & 1406 (RT) & \cite{WilliamsBook, Wang94} \\
  &   &  - & 1462 (RT) & 1416 (RT)& \cite{Wang94} \\
				        &   &  - & 1445 (RT) & 1415 (RT) & \cite{Yamamoto05} \\
(d$_8$-ET)AuBr$_2$Cl$_2$ & - & - & 1445 (RT) & 1415 (RT) & \cite{Yamamoto05} \\
($^{13}$C-ET)$_2$AuBr$_2$Cl$_2$ & - & -  & 1445 (RT) & 1355 (RT) & \cite{Yamamoto05} \\
(ET)(ClO$_4$) & - & - & 1445 (RT) & 1415 (RT) & \cite{Yamamoto05} \\
\hline
\textbf{1:2 salts}\\
(ET)(ClO$_4$)$_2$ & - & - & 1303 (RT) & 1325, 1376 (RT) & \cite{Wang96} \\
\hline
\textbf{BETS based systems}\\
\textbf{(2:1 salts)}\\
$\alpha$-(BETS)$_2$I$_3$ & \textit{P}$_1$&  1448,1461 (300) & - & - & \cite{Priya25} \\
 		 & 			 &  1436, 1443, 1453,  & - & - & \cite{Priya25}  \\
 		 & 			 &  1466, 1471, 1477 (10)  &   &   & 			 \\
$\kappa$-(BETS)$_2$Mn[N(CN)$_2$]$_3$ & \textit{P}2$_1$/c &  1456 (300) & - & - & \cite{Schmidt24} \\
   & - &  1459, 1462 (12) & - & - & \cite{Schmidt24} \\
$\kappa$-(BETS)$_2$GaCl$_4$ & \textit{Pnma} & 1448, 1454 (300) & - & - & \cite{Iakutkina21b} \\
$\lambda$-(BETS)$_2$GaCl$_4$ & \textit{P}$\overline{1}$ &  1454, 1458 (300) & - & - & \cite{Iakutkina21b} \\
\hline\hline
\end{longtable}
$^*$: $\nu_2$ and $\nu_3$ modes are coupled for these $\theta$-phase compounds\\

\pagebreak

\twocolumngrid

\appendix

\section{Temperature dependence of vibrational modes}

\label{sec:TemperatureDependence}
\begin{figure}[h]
    \centering
        \includegraphics[width=0.6\columnwidth]{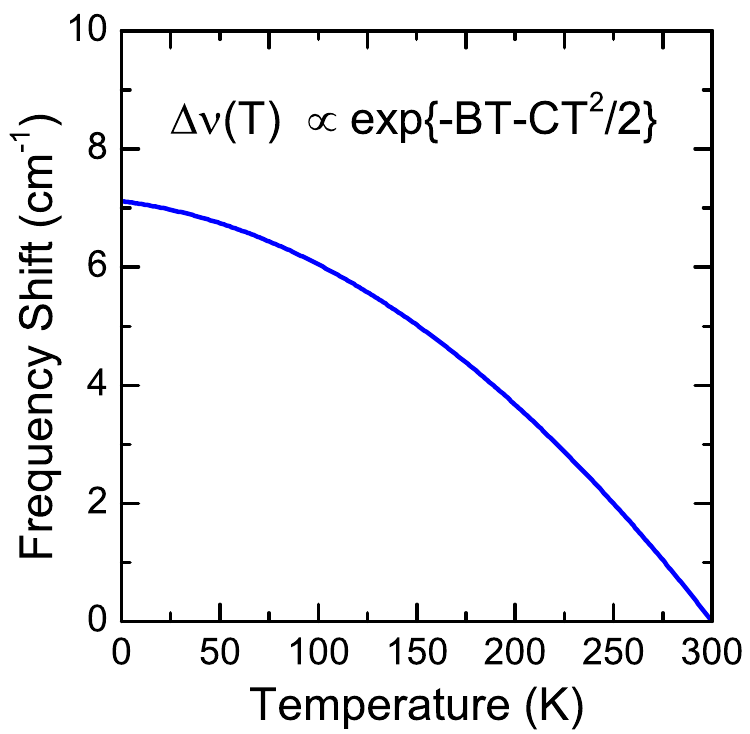}
        \caption{Shift of the vibrational frequency as a function of temperature compared to the room temperature value. The line corresponds to Eq.~(\ref{eq:temperaturedependence}) with $\nu_0 = 1500$~\cm, $B=2.8 \times 10^{-6}~{\rm K}^{-1}$ and $C= 8.7 \times 10^{-8}~{\rm K}^{-2}$, as example.}
    \label{fig:temperatureshift}
\end{figure}

The thermal contraction the crystal lattice leads to an up-shift of the phonon resonance frequency because
the increased charge density at the bonds hardens the modes. The dependence of the phonon frequency $\nu(T)$ on temperature is generally described by
\begin{equation}
\frac{1}{\nu} \frac{{\rm d}\nu}{{\rm d}T} = -\gamma \beta(T) \quad ,
\label{eq:Gruneisen}
\end{equation}
where $\gamma$ is the optical Gr{\"u}neisen parameter and $\beta(T)$ is the coefficient of volume thermal expansion, which in first order is $\beta(T) = \beta_0 +\beta_1 T$. In the case of molecular solids, in addition to lattice phonons molecular vibrations are present mainly in the mid-infrared region. At lower frequencies both internal and lattice vibrations are coupled and cannot be distinguished \cite{Dressel92,Liu97,Demiralp97,Girlando00,Dressel16}.
When cooling from $T=300$ to 10~K , the vibrational features typically harden by 5-7~\cm;
this blue-shift can be expressed by:
\begin{equation}
\nu (T) = \nu_0 \exp\left\{ - B T - C T^2/2 \right\}
\quad .
\label{eq:temperaturedependence}
\end{equation}
For the examples of \kCuCl\ and $\beta'$-(BEDT-TTF)$_2$ICl$_2$, given by Yakushi \cite{Yakushi12}, the temperature behavior of the $\nu_2$ modes is well described
by
$B = \gamma \beta_0 \approx 2.8 \times 10^{-6}~{\rm K}^{-1}$ and $2.7 \times 10^{-7}~{\rm K}^{-1}$,
$C = \gamma \beta_1 \approx 8.7 \times 10^{-8}~{\rm K}^{-2}$ and
$8.1 \times 10^{-8}~{\rm K}^{-2}$ , respectively.
In Fig.~\ref{fig:temperatureshift} the temperature dependence of the vibrational frequency is plotted. For simplicity, we assume $\nu_0 = 1500$~\cm, which yields a room-temperature value of 1493~\cm\ and a blue-shift $\Delta \nu$ of about 7~\cm\ at low temperatures.

The hardening of the other C=C stretching modes appears similar; hence we suggest the same relative frequency shift for the $\nu_2(a_g)$ and $\nu_{27}(b_{1u})$ vibrations.
For our comparison of spectral data recorded at different temperatures, the mode frequencies are adjusted according to Eq.~(\ref{eq:temperaturedependence}).

\bibliography{ref}

\end{document}